# Spatiotemporal attosecond control of electron pulses via subluminal terahertz waveforms

MIKHAIL VOLKOV*, ERUTHUPARNA RAMACHANDRAN, MAXIMILIAN MATTES, ATAL BIHARI SWAIN, MAXIM TSAREV AND PETER BAUM*

*Universität Konstanz, Universitätsstraße 10, 78464 Konstanz, Germany*
*mikhail.volkov@uni-konstanz.de, *peter.baum@uni-konstanz.de*

**Abstract:**

Shaping electron beams with the cycles of light provides femtosecond and attosecond time resolution in electron microscopy and enables fundamental quantum-coherent measurements. However, efficient light-electron control requires a prolonged interaction between the two beams for cascaded transfer of photon energy and momentum to the freely propagating electrons. Here we report the use of traveling evanescent terahertz waves to achieve velocity matching and thereby high acceleration gradients both in space and in time. With experiment and simulations, we demonstrate attosecond streaking, temporal pulse compression, acceleration and spatial focusing of sub-relativistic electron pulses with a single evanescent-wave element under the control of selected terahertz delays and phases. Based on these results, we propose to use a symmetric arrangement with two evanescent terahertz waves to generate isolated attosecond electron pulses in a beam with realistic parameters. These results establish subluminal terahertz waves as a promising tool for ultrafast electron pulse control.

## 1. Introduction

The all-optical manipulation of free-electron beams with the optical cycles of coherent light [1–3] has enabled substantial advances in ultrafast electron microscopy and free-electron quantum optics [4–9]. For example, laser control of an electron beam has provided femtosecond [1,10,11] and attosecond [12–14] time resolution and opened up novel regimes for quantum measurements [15,16].

In free space, electrons and photons cannot easily exchange energy due to the need for momentum conservation. In all the above experiments, light-electron coupling is therefore achieved with help of a third body as a modulation element, such as a field-enhancing micro/nano-structure [5,17,18], a thin foil [12–14,19] or an evanescent wave [20–22]. The effect on the electrons is then typically analyzed with an electron energy loss spectrometer (EELS) and real-space images can be obtained via photon-induced near-field microscopy (PINEM) [4]. However, since electron energy analysis alone is insensitive to the temporal structure of the shaped electron pulses, a second light-electron interaction is needed for a complete pulse characterization in time [1,8,12,23]. This is achieved, for example, with angular streaking [1,19] or by adding a second time-dependent momentum change, followed by energy analysis [12,14]. A well-defined electric field, as well as an efficient electron streaking method are therefore essential for unambiguous high-resolution results.

A central parameter for light-electron modulation is the laser wavelength [24]. While visible or infrared laser light provides a short cycle period that is useful for shortest-pulse generation [12,13,25] terahertz radiation offers a much higher ponderomotive potential and a well-defined carrier-envelope phase. Due to the long wavelength, it can cover an incoming electron pulse entirely in time. Also, it can provide rather homogeneous acceleration gradients for efficient manipulation and characterization [26] of large-diameter electron beams with low emittance. Terahertz radiation have for example been used for linear electron acceleration [3,11], streaking [3,27–29] and temporal pulse compression [1,3]. Significant enhancement of coupling strength for electron acceleration can be achieved, for example, by matching the electron velocity with the phase velocity of the terahertz wave using waveguides [2], however, it requires radially polarized terahertz pulses and restricts the transverse dimensions of the electron beam. Recently, it was proposed theoretically to use evanescent terahertz waves for particle acceleration [30–32]. The idea is to use the refractive index of a dielectric material to slow down the phase velocity of the THz wave such that it matches the group velocity of the electrons. However, the link between the longitudinal and transversal acceleration gradients via the Panofsky-Wenzel theorem [33] can induce spatial chirp and thus challenges the spatial homogeneity of ultrafast electron pulse manipulations and in particular impedes the ability for sub-femtosecond temporal compression or attosecond streaking. Consequently, research in terahertz control of electron beams must strive at combining highest possible interaction lengths with lowest possible distortions, in order to create the desired effects without undesired phase space manipulations [34].

## 2. Idea and experimental setup

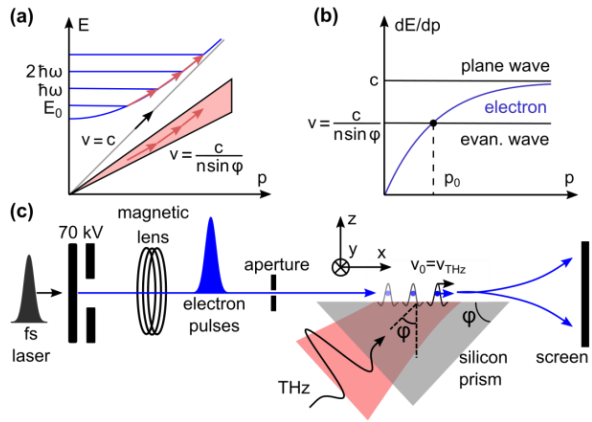

Fig. 1. (a) Energy-momentum diagram showing the dispersion of a free electron (blue curve), a free photon (gray line) and a traveling evanescent wave in a focused beam (red-shaded area). Here, n is the refractive index of a dielectric and φ is the incidence angle of the beam that reflects at dielectric-vacuum interface and produces the evanescent wave. A range of wavevectors (red arrows) enables cascaded processes by energy-momentum matching along the dispersion curve. (b) Velocity matching diagram in (v = dE/dp, p) axes. (c) Experimental setup. Electron pulses at 70 keV central energy and 50 fs duration are focused with a magnetic lens and filtered with a 50-µm aperture. Silicon prism with an angle φ = 38° provides velocity matching of the Terahertz evanescent field with the propagating electron pulses.

Here, we demonstrate the applicability and use of velocity-matched evanescent Terahertz waves for attosecond electron streaking, space-time electron bunch characterization, and temporal compression. We produce a traveling evanescent terahertz wave by illuminating a silicon-vacuum interface and use the evanescent wave to control an electron bunch near the surface. The presence of various phase-locked longitudinal and transversal fields in the emerging wave enable versatile manipulations of electron trajectories. By experiment and simulations, we demonstrate streaking, acceleration and temporal compression of electron pulses by simply tuning the electron-terahertz delay. Furthermore, a symmetric scheme with two evanescent waves offers a simple way to produce isolated attosecond electron pulses at a desired distance.

In free space, electron and a single photon with an energy much smaller than the electron rest mass, cannot directly interact. Figure 1a illustrates the necessary energy-momentum conservation. Since the free-electron dispersion curve (parabola in Fig. 1a) never crosses the light line (grey line in Fig. 1a), the electron momentum increment $dp = dE/v$ (where $v$ is the electron velocity) is always larger than the photon momentum $dE/c$ (Fig. 1b). To reduce the phase velocity of light, we use an evanescent wave that emerges above a surface of a dielectric material when a beam of light illuminates the dielectric-vacuum interface from the inside (total internal reflection). Thereby, the momentum and energy increments can be matched (intersection with the blue curve in Fig. 1b), enabling efficient interaction with electrons that travel along the surface in free space. By changing the incident angle $\varphi$ of the optical beam, the velocity matching can be tuned to particular electron energies. Importantly, this momentum matching is relaxed, because a focused light beam provides a wavevector spread in the evanescent wave that enables broadband velocity matching and surfing of the electron for an extended time (red-shaded area and red arrows in Fig. 1a). The electric and magnetic field vectors in an evanescent wave can point in various longitudinal and transverse directions in dependence on the incoming polarization and phase delay (see e.g. [20]). An evanescent wave at terahertz frequencies [30] should therefore be a versatile and effective way for controlling an electron beam in space and time.

Figure 1c depicts the experiment. We use the second harmonic of 1030 nm, 270-fs laser pulses from a chirped-pulse laser amplifier (Pharos, LightConversion) to produce single-electron pulses [35] via two-photon photoemission from a gold-coated photocathode [36]. All pulse durations and beam widths are specified as full width at half maximum (FWHM). The laser-emitted electron pulses are accelerated to a kinetic energy of 70 keV and then pre-compressed to a duration of 50 fs with Terahertz radiation on a planar metal membrane (see Ref. [10] for details). These electron pulses then propagate along a 20-mm-long surface of a silicon prism with an apex angle of $\varphi = 38°$ (Fig. 1c). The two Terahertz pulses, the one for electron pre-compression on the membrane [10] and the one in the prism, are produced via Cherenkov mechanism in a $LiNbO_3$ crystal [37]. The second Terahertz beam is focused into the silicon prism with a 45° off-axis parabolic mirror (focal length of 59 mm) with perpendicular incidence into one of the prism facets; see Fig. 1(c). Inside the prism, the beam undergoes a total internal reflection at the silicon-vacuum interface. The Terahertz pulses inside the prism are single-cycle waveforms with a central frequency of 0.3 THz and an estimated pulse energy of 0.7 nJ. The resulting beam waist at the surface is ~3 mm in $x$ direction (the direction of velocity matching) and ~2 mm in $y$ direction (perpendicular to the electron beam). At the chosen angles of the prism, the phase velocity of the evanescent wave in the direction of the electron propagation matches to the electron's group velocity:

$$v_{THz} = \frac{c}{n}\frac{1}{sin\varphi} \approx v_0 \approx 0.48\, c, \tag{1}$$

where $v_{THz}$ is the phase velocity of the evanescent wave $n=3.418$ is the refractive index of silicon [38]. This velocity matching is very broadband: for the chosen geometry and finite Terahertz beam diameter, an electron energy mismatch of ±5 keV around the 70-keV central energy results in an accumulated phase slip of less than 10% of one optical cycle.

Under this velocity-matching condition, the interaction of the evanescent wave with electrons can be analytically derived. The dimensionless electron-photon coupling strength is [22,39]:

$$g = \frac{\sqrt{\pi}w_0 eE}{\hbar\omega} \tag{2}$$

where $w_0$ is the Gaussian beam waist. We notice that $g$ is proportional to $E(z,t)w_0$, that is, the product of electric field strength (determining the force) and beam waist (determining the surfing length). For example, assuming a 70-keV electron beam a diffraction-limited Terahertz focus with a size of $\sqrt{\pi}w_0 \approx \lambda \approx 1$ mm with 1 MV/m peak field strength results in keV-level electron acceleration and corresponds to $g \sim 10^6$. In the angular streaking regime, we can expect a maximum deflection angle $\theta = \Delta v_\perp/v_0$ of ~25 mrad, and the streaking speed is ~150 μrad/fs.

Effectively, the presented geometry offers three central features: First, the electron does not need to pass through any kind of material that could spoil the electron beam quality or the spatiotemporal coherence of the wave function. Second, the interaction is velocity-matched, providing efficient light-electron coupling. Third, evanescent waves are stable and well-defined waveforms that do not alter the temporal shape of the incoming terahertz pulses. In contrast to micro-structured resonators [40] or segmented terahertz electron accelerator and manipulator STEAM devices [3], the field shape of a single-cycle waveform is therefore directly imprinted on the electron beam without a limited bandwidth, similar to metal membranes [1] but at a substantially better efficiency.

## 3. Results

### A. Evanescent field characterization

In a first set of calculations and experiments, we consider the electromagnetic field vectors of the evanescent Terahertz waveform. We assume a p-polarized sine-type single-cycle terahertz pulse with a duration of 1.6 ps [37], focused into a 3-mm spot at a pulse energy of 0.7 nJ. These parameters produce an electric-field amplitude of 0.3 MV/m inside the prism. Figure 2a depicts the calculated electric field vectors [20] at a selected moment in time. Figure 2b shows the corresponding $E_x$ and $E_z$ components at the pulse centre at the prism surface. The $E_z$ component can mediate a sideway deflection of the electrons while $E_x$ can provide a longitudinal acceleration. The magnetic component of the evanescent wave is perpendicular to $E_z$ and the Lorentz force on a travelling electron is therefore antiparallel to $-|e|E_z$, reducing the deflection but not effecting acceleration/compression (see Ref. [34] and the Appendix).

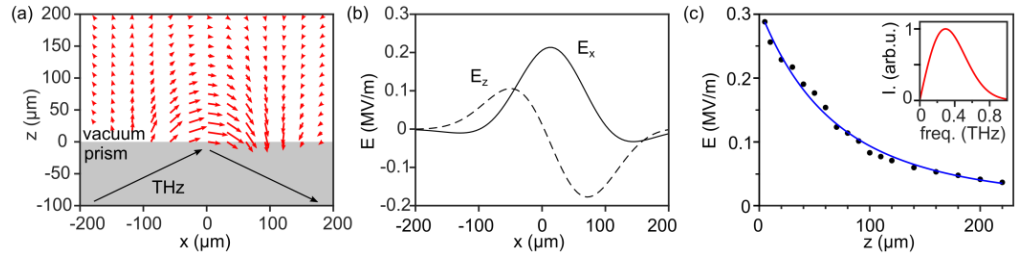

Fig. 2. Field vectors and depth profile of an evanescent terahertz wave that emerges out of a dielectric prism surface. (a) Time-frozen snapshot of the electric field. (b) Components $E_x$ and $E_z$ of the electric field vector just above the prism surface. (c) Measured (black dots) and fitted (blue curve) evanescent wave amplitude in free space. The effective 1/e-attenuation length is 74 μm. Inset: amplitude spectrum of the terahertz pulses.

In order to confirm this shape of the evanescent terahertz wave by experiment, we measure the position-dependent deflection of the electron beam as a function of distance to the surface ($z$ axis). The prism position and the Terahertz alignment are kept constant during this experiment. The deflection is proportional to the field strength via Eq. 4 (see Appendix). Figure 2c shows the evaluated evanescent field amplitude as the black dots, derived from the measured deflection angles. We see a decay on a length scale of ~100 μm. Since the evanescent field stretches further out for longer wavelengths, the measured field amplitude fades non-exponentially (blue line). Assuming a terahertz amplitude spectrum in the form of $\omega \exp(-\omega^2\tau^2/16ln2)$, where $\tau$ is the effective terahertz pulse duration and $\omega$ is the angular frequency [37,41], we calculate the spectrally-weighted evanescent field amplitude according to $exp(-ikz)$. The blue line in Fig. 2c shows the best fit, yielding $\tau = 1.6$ ps and a spectral maximum at $\omega = 0.3$ THz (Fig. 2c, inset), in agreement with the expectation from our Cherenkov generation scheme [37]. The measured electron beam deflections (dots) fit well to the analytical deflection model (see section C), confirming the validity of the evanescent-wave assumptions in the calculations as well as the presence of the intended velocity matching.

### B. Streaking and pulse reconstruction

Modern technology offers several ways for compressing electron pulses to unprecedentedly short duration [12,13,25], but all-optical pulse characterization [1] is often not directly available on the same level of time resolution. In this section, we therefore consider the use of our terahertz-illuminated prism for the temporal characterization of isolated ultrashort electron pulses with few-femtosecond and attosecond time resolution. In particular, we show that the terahertz waveform as well as the time-dependent charge density of the electron beam together with some of its spatiotemporal coupling coefficients [10] can be reconstructed from a single electron deflectogram.

The principle of our pulse characterization is an all-optical streak camera via time-dependent beam deflection [40]. Critical for the ability to measure extremely short electron pulses, for example few-femtosecond [10,42,43] or attosecond pulses [12,13,25], is the streaking speed, that is, the amount of deflection angle change per unit of time. So far, in the pursuit for a highest terahertz streaking gradients, several field-enhancing structures have been proposed. A slit structure [27], a parallel plate waveguide [28], a split-ring resonator [29] and tailored segmented terahertz electron accelerator and manipulator (STEAM) devices [3] provide angular streaking speeds of 5 µrad/fs, 7 µrad/fs, 4 µrad/fs and 140 µrad/fs, respectively, yielding a sub-10-fs resolution. Circularly polarized pulses increase the time resolution by streaking in two dimensions [44]. Attosecond streaking of free electrons with infrared light was recently demonstrated with a streaking speed of 100 µrad/fs on a planar dielectric membrane [45].

In the experiment, we arrange the electron beam rather close to the surface, but without losing intensity by clipping. Under these conditions, the center of the electron beam is ~20 µm away from the surface. We record deflected electron beam profiles as a function of the terahertz delay, integrating over the redundant coordinate ($y$ axis). Figure 3a shows the resultant deflectogram. We observe a streaking slope of 250 as/pixel, corresponding to an angular streaking rate of 46 µrad/fs. The un-streaked electron beam diameter is 4 pixels or 62 µm on the screen.

We compare the experiment with classical-trajectory Monte Carlo calculations in Fig. 3b. In the calculations we assume normal distributions of electrons with a beam diameter of 50 µm, and an angular beam spread of 0.5 mrad. These values correspond to an emittance of 10 nm·rad, typical for electron diffraction beamlines [46] but ~1000 times worse than in modern TEMs. The energy bandwidth is set to 1.3 eV [10]. The temporal shapes of the terahertz and electron pulses are then considered free fit parameters. The best match with the experiment is achieved by assuming a Gaussian electron pulse of 50 fs duration, and a 280-fs background with temporal skewness of $\gamma = -5$ (defined as the temporal distribution's normalized third moment) and a Terahertz phase offset $\varphi = -0.31$ rad with respect to a cosine pulse. We find an excellent agreement with the experiment, including the fine features of the deflectogram (compare Fig. 3b with Fig. 3c).

Several features stand out in the results of Fig. 3b. First, the cycle sequence of the terahertz pulses is not broadened by resonances or distortions and the temporal waveform therefore directly exerts its forces onto the electrons without distortions in the time domain. There are only small ripples in the region 1.5-2.5 ps, that stem from residual reflection from the prism edges. The two turning points of the deflectogram have slightly different absolute amplitudes, indicating that the incoming terahertz pulse has a small carrier-envelope phase offset from the cosine shape that is expected from the generation [37]. The group velocity and phase velocity of silicon at sub-THz frequencies are very similar; therefore, we assign this observation to the role of Gouy phase shifts along the focus direction.

Second, the streaking speed around t = 0 is 400 as/pixel at a terahertz peak field strength of only 0.3 MV/m. By moving the electron beam closer to the prism, we obtain an even faster streaking slope of 250 as/pixel or 46 µrad/fs (Fig. 3b, inset), at the expense of partially cropping the electron beam on the prism at negative delays. This streaking speed is about three times lower than in a STEAM device [3], but our electron beam size is 6 times smaller, providing therefore a better, 1-fs time resolution. Crucially, our terahertz pulses are also ~1000 times weaker. Normalized to terahertz pulse energy, the present prism schematic is therefore more efficient. The reason for the giant streaking effect is the velocity-matched cycle-surfing of the electrons on the terahertz wave during the entire travel time along the prism surface.

Third, we see in Fig. 3b around a delay of -1 ps an asymmetry in the delay-dependent intensity distribution. Deflected electrons appear below and to the right of the main trace in this region, but not above and to the left. The feature shows the presence of electron pulses with an asymmetric shape in time. We associate this profile with inelastic scattering background from the pre-compression membrane [10]. Electron with a few eV of energy loss will arrive a few fs later at the prism [47] and therefore produce an electron pulse with a slight tail. The deflectogram reveals that the current density of the tail is <20% of the main pulse.

Fourth, we see at the positive and negative turning points of the data at $t \approx -0.5$ ps and $t \approx 0.5$ ps a substantial spread along the deflection direction, although the terahertz field and therefore the deflection is constant in time. We explain this observation by the effects of the finite spatial extent of the electron beam in our experiment, which is roughly a half of the depth of the evanescent wave (compare Fig. 2c). Electrons closer to or farther from the surface therefore undergo different amounts of deflection. Consequently, the deflectogram broadens in angular direction, even for electron pulses of infinitely short duration. This effect is directly proportional to the mean deflection and therefore most prominent at the turning points of the deflectogram.

To emphasize the direct sensitivity of prism streaking to all of these features, we report in Fig. 3d a simulation in which we assume a background-free time-symmetric electron pulse, a ray-like electron beam ($d = 0.1$ µm) and a terahertz waveform without carrier-envelope phase offset. The delay-dependent intensity distribution is now always symmetric around its maximum, because the electron pulse has simply a Gaussian shape. The asymmetry between the maximum positive and maximum negative deflection are now equal, because the Terahertz waveform is cosine-like. The intensity of the deflectogram is higher at the turning points (Fig. 3d, inset) as compared to $t = 0$, because the electron pulse with its finite duration is not streaked and dispersed at these times.

In summary, we see that the temporal distribution of ultrashort electron pulses can be well characterized at ultimate time resolution by exposing them to the spatial and temporal gradients of our evanescent-wave streaking arrangement. Both the Terahertz waveform and the electron charge density are almost directly deducible from the raw data without a reconstruction algorithm. In case that a more automatic and complete reconstruction is desired, phase-retrieval methods can be applied, for example with a principal component generalized projection algorithm (PCGPA) that is routinely used in attosecond photoelectron streaking to characterize pump and probe pulses simultaneously [48].

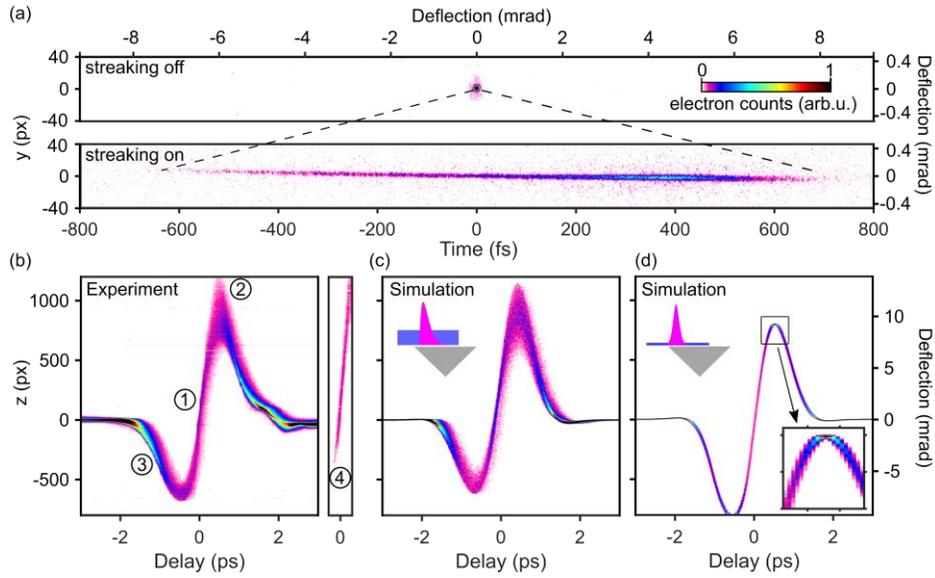

Fig. 3. Streaking traces. (a) Electron beam as detected on the camera with streaking off (upper panel) and streaking on (lower panel). (b) Measured deflectogram (streaking as a function of time delay). The streaking slope is 400 as/pixel (1). The turning points (2) are blurred by residual inhomogeneities of the evanescent wave for a thick electron beam. Slices at the increasing slow (3) reveal a temporal asymmetry (skewness) of the charge density of the electron pulses. The inset (4) shows the steepest streaking slope of 250 as/pixel or 46 µrad/fs, obtained by moving the electron beam closer to the prism. (c) Simulated deflectogram. All features are reproduced. (d) Numerical simulation for a narrow electron beam with symmetric temporal shape and zero carrier-envelope phase offset.

**C. Focusing, acceleration and compression**

For the purpose of electron acceleration and control in time, researchers have come up with a great variety of interaction elements so far. For example, planar membranes and field-enhancing structures enable electron temporal compression at terahertz, mid-infrared and optical frequencies [1,10,12,13]. An intrinsic limitation of these schemes is their finite interaction length, which is typically much shorter than one optical wavelength. This issue can be overcome by using a tailored waveguide [49], pulse-front tilt [50], periodic dielectric structures [51,52] or near-infrared evanescent fields in what has been called an inverse Cherenkov scheme [20–22].

In this section, we show by numerical simulations that the reported velocity-matched electron-terahertz interaction with prism surfaces is a nearly all-purpose electron manipulation device that can be used, for example, for efficient electron acceleration, electron pulse compression or spatial focusing, similarly to a STEAM device [3]. While the latter device usually operates at µJ-level terahertz pulse energy, our scheme uses sub-nanojoule pulse energies but nevertheless yields almost equal levels of electron pulse manipulation at a comparable initial electron energy (70 keV). Crucially, this feature allows one in practice to work with terahertz pulses at much higher laser repetition rates, up to the MHz regime, and thereby obtain superior signal-to-noise ratios in electron diffraction and microscopy experiments.

Basis of the following arguments is the demonstrated agreement between our measurement results and the numerical simulations (Figs. 2,3). In the subsequent simulations, we apply the same geometry as above but now consider four different carrier-envelope phases of the terahertz waveform and their resulting electron distributions in space, time and energy after the interaction. As above, we assume a 50-um-wide electron beam with a z-offset of 20 µm and a pulse duration of 50 fs. The terahertz pulse has again a peak field strength of 0.3 MV/m, a pulse duration of t = 1.6 ps, a beam waist of 3 mm and a spectral maximum at 0.3 THz. The electrons in the pulse are assumed to be point particles that travel in the simulated electromagnetic field of the evanescent wave at a fixed electron-terahertz delay.

In a first example, we look in more detail into the above-explained temporal streaking geometry, that is, the effect of our prism device for a sine-like terahertz waveform into which the electron pulses are injected around the zero-crossing of the waveform. Figure 4a shows the position-dependent electron trajectories. We see that the electron beam goes through almost without distortions. Figure 4b shows the initial time-position distribution of the incoming electrons (black) and the resulting spatiotemporal distribution after the interaction (violet). As expected, the front electrons are deflected upwards and the tail are deflected downwards by the time-dependent $E_z$ field. The $E_x$ field always decelerates the electrons, with the front and tail electrons decelerated less than the center electrons. In combination, the effects of the $E_z$ and $E_x$ field components produce tilted electron pulses at a distance of 9 mm from the terahertz focus. This coupling of longitudinal and transverse acceleration, also evident from the fields depicted in Fig. 2a, is a consequence of the Panofsky-Wenzel theorem [33], which states that the transverse momentum changes only when longitudinal acceleration depends on the transverse coordinate.

In a second example, we apply a cosine-type terahertz pulse. Figure 4c shows the trajectories and Fig. 4d the spatiotemporal distribution before (black) and after (blue) the interaction. At this carrier-envelope phase, the prism produces an upwards deflection of the entire electron beam that is accompanied by a spatial focus at a distance of around 9 mm. The reason for this latter phenomenon is the lensing effect of the evanescent-wave. Simultaneously, the initial electron pulse duration is broadened (Fig. 4d, blue vs black dots) by the time-dependent $E_x$ component of the field, which accelerates the front of the electron pulse and decelerates its tail.

In a third example, we consider an inverted cosine pulse (Fig. 4e). We start here with an electron beam that is shifted slightly more away from the prism surface (75 μm) than in the other examples (20 μm). We see a downwards deflection of the entire electron beam that is accompanied by a defocussing effect. The origin of this phenomenon is the negative lensing effect of the evanescent wave. Figure 4f shows the arrival time distortion of the electrons before the interaction (gray) and after the interaction at a distance of x = 20 mm (magenta). We see that the initial 50-fs long electron pulse is compressed down to a duration of 2.5 fs. Figure 4h shows the electron energy spectrum before (black) and after (magenta) the interaction. We see a broadening from 1.3 eV to 60 eV. Such a spectral broadening is expected when compressing pulses due to the conservation of phase space volume.

In a fourth example, shown in Fig. 4g, we apply an inverted sine pulse with a carrier-envelope phase that is shifted by $\pi$ with respect to the waveform in Fig. 4a. In this geometry, there is a strong longitudinal acceleration by the $E_x$ component of the evanescent wave. Consequently, we see that the initial electrons energy spectrum at 70 keV (black) shifts after the interaction to a new centre at around 71.25 keV (violet). At the same time, the spectrum broadens to a width of ~400 eV due to position-dependent electron acceleration in the evanescent wave.

These four examples and the detailed simulations reveal that our prism geometry with its sub-luminal evanescent wave can act as a versatile device for electron pulse compression, streaking, acceleration and time-dependent focusing, although each of these operational modes is unavoidably accompanied by secondary effects. For example, streaking implies changes of central energy and generation of tilted pulses, pulse compression implies spatial defocussing and electron acceleration implies beam deflection and broadening of the spectrum. Depending on the application, these secondary effects can be detrimental, but a smaller electrons beam diameter and/or shorter initial pulses reduce the distortions, typically with inverse linear proportionality.

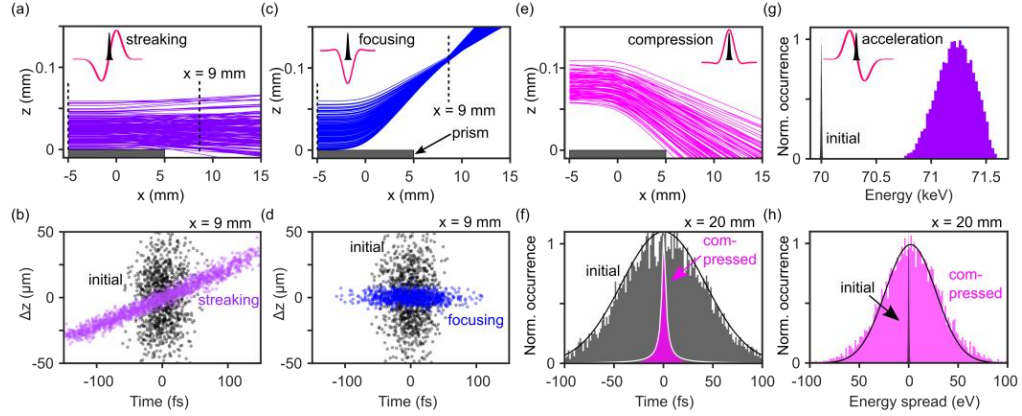

Fig. 4. THz-phase controlled electron manipulation. (a) Trajectories for an electron pulse injected at the terahertz $E_z$ slope for streaking. (b) Spatiotemporal distributions of the initial (black) and streaked (violet) electrons. (c) Election pulse injected at the maximum of the terahertz $E_z$ field causing spatial focusing (d) Corresponding spatiotemporal electron distributions. (e) Same as panel (b), but with the $E_z$ component in the opposite direction, causing defocusing and simultaneous compression. (f) Pulse duration as histograms before (black) and after compression (magenta), with a Gaussian fit of 50 fs and a quasi-Lorentzian fit with FWHM of 2.5 fs. (g) Electron energy distribution before (black) and after acceleration (violet) by a minus-sine THz pulse. (h) Initial (black) and compressed (magenta) electron energy distributions, with Gaussian fits of 1.3 eV and 60 eV, respectively.

## D. Double-prism geometry for attosecond electron pulse compression

Based on the above results, we now consider a symmetric version of our prism geometry for the purpose of electron pulse compression towards attosecond duration at minimum distortions. Such an arrangement has been previously suggested for proton acceleration [30]. Figure 5a depicts the geometry. Two identical dielectric prisms (grey) are placed symmetrically with respect to the optical axis of the electron beam (magenta). The evanescent waves from each prism constructively interferes in the gap, creating a nearly homogenous field distribution in the middle of the gap.

In the simulations, the electron beam has the parameters of a typical transmission electron microscope, central energy of 200 keV, energy spread of 0.8 eV, emittance of 5 pm·rad and beam diameter of 1 µm. The electron pulse duration is 300 fs. For the optical geometry, we assume a pair of 4-mm-long silicon prisms with a gap of 100 µm. The terahertz pulses have an amplitude of 1 MV/m and cosine-shaped waveform. We use analytical expressions for the terahertz field in the gap between the prisms obtained as the exact solution of the Maxwell equations for Gaussian beams in a three-layer infinite media.

Figure 5b shows the electric field vectors created by the two evanescent terahertz waves at time zero. We see a rather a homogenous field that should allow to minimize the longitudinal-transverse coupling and produce electron compression almost without secondary distortions. Figure 5c shows the results of the Monte Carlo trajectory simulations. The magenta histogram shows the time distribution of the incoming electron pulses and the black data shows the compressed pulses. Figure 5d shows a zoom into this data: We see a pulse shape with a pulse duration of about 200 as. The schematic therefore provides more than a thousand-fold compression over just 10 mm of propagation. The slightly non-Gaussian shape close to the base line stems from residual nonlinearities ion the time-dependent acceleration gradient; compare the emergence of a background current in multi-cycle attosecond experiments [12,13]. Figure 5e shows the position-dependent charge density as a function of time. We see that most of the charge density is concentrated in a 8-µm beam, with no pulse tilt. Spatiotemporal distortions are therefore not a limiting factor of our reported geometry. The limitations remain the quantum dynamics of the compression [53], the initial electron energy spread and the finite emittance and beam diameter of the electron beam.

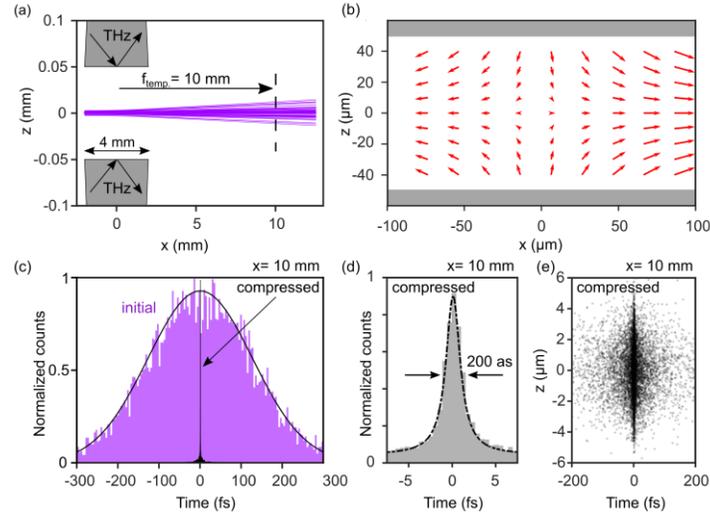

Fig. 5. Proposed symmetric THz compressor for direct production of attosecond pulses. (a) Two prisms (grey) and the electron trajectories (violet). (b) The electric field vector map (red arrows) between the prisms (grey). (c) Electron pulses before (violet) compression, and after compression at the temporal focus (black). (d) The compressed pulse shape, with a duration of 200 as. (e) Charge density distribution of the compressed electron in space and time.

## 4. Conclusion and Outlook

We demonstrated experimentally that single-cycle THz waveforms of sub-nanojoule energy can produce attosecond streaking gradients for the free electrons via prism-enabled velocity-matched interaction. The longitudinal and transversal field gradients of the evanescent wave allow to fully characterize the electron pulses in time and space. Calculations show that the same prism structure can be used to streak, focus, accelerate and compress the electron pulses. In experiment, we reach time resolution of 1 fs in the electron beamline. Furthermore, our simulations predict that a symmetric layout with two prisms will allow to minimize the longitudinal-transverse coupling and thus selectively utilize compression. The proposed symmetric scheme has the potential to deliver isolated electron pulses of attosecond duration in a transmission electron microscope.

**Funding.** This project has received funding from the European Union's Horizon 2020 research and innovation programme under the Marie Skłodowska-Curie grant agreement No 896148-STMICRO. Atal Bihari Swain acknowledges support from the Alexander von Humboldt Foundation.

**Disclosures.** The authors declare no conflicts of interest.

**Data availability.** The code for the classical-trajectory Monte-Carlo calculations is available at https://github.com/mikhailVvolkov/InverseCherenkovEffect as a GUI application.

**Appendix: Emittance and time resolution in streaking.**

Here we calculate the influence of beam emittance on the streaking temporal resolution. Given divergence $\sigma$ and angular streaking speed $d\theta/dt$, the attainable temporal resolution is:
$$\Delta t = \sigma \cdot (d\theta/dt)^{-1}. \qquad (3)$$
From Newton's equation with Lorentz force, under velocity matching, the deflection angle $\theta = p_\perp/p$ can be approximated as:
$$\theta = \frac{eE(z,t)}{p}\left(1 - \frac{v}{cn\sin\varphi}\right)\frac{\sqrt{\pi}w_0}{v} = \frac{g\hbar\omega}{pv\gamma^2} = \frac{g\hbar\omega}{mv^2\gamma^3} \qquad (4)$$
where $p$ and $v$ are the electron momentum and velocity, and the bracket term (equivalently $1/\gamma^2$) corresponds to the magnetic field effect that reduces the transversal streaking.

From Eq. (4), the angular streaking speed is $\omega\theta$, whereas the lensing effect $d\theta/dz = k_{evan}\theta$ produces an angular distribution $\sigma = k_{evan}\theta d$, where $d$ is the electron beam diameter and $k_{evan} = \omega/(v\gamma)$ is the velocity-matched evanescent wave decay parameter. The best achievable temporal resolution $\Delta t = d/(v\gamma)$ is therefore determined solely by the electron beam transverse dimension. Consequently, sub-cycle resolution $\Delta t < 1/\omega$ sets a constraint on the electron beam size $d < \lambda$, and therefore on the source emittance:
$$\varepsilon < \theta\lambda, \qquad (5)$$
where we use the non-normalized emittance definition $\varepsilon = d \cdot \sigma$. Therefore, sub-cycle characterization of electron beam with low emittance requires high field amplitude. A characteristic deflection angle of 1 mrad therefore translates into emittance of $\varepsilon < 1\mu m \cdot rad$ for THz frequency and $\varepsilon < 1nm \cdot rad$ for visible light. The condition for THz streaking is satisfied in an electron beamline [46], whereas the condition for the visible light is easily fulfilled in a transmission electron microscope (TEM) which typically has an emittance of 5 pm·rad. Thus, the device can be used for sub-fs resolution streaking in the microscope and few-fs resolution in the beamline.